\documentclass{article}
\usepackage{spconf}
\usepackage{amsmath,amssymb,amsfonts}
\usepackage{algorithmic,booktabs}
\usepackage{graphicx,subfig}
\usepackage{textcomp}
\usepackage{xcolor}
\usepackage{lipsum}
\usepackage{xurl}

\title{LiDAR Depth Map Guided Image Compression Model}

\name{\begin{tabular}{c}Alessandro Gnutti$^\star$, Stefano Della Fiore$^\dagger$, Mattia Savardi$^\mathsection$\\ Yi-Hsin Chen$^\ddagger$, Riccardo Leonardi$^\star$, and Wen-Hsiao Peng$^\ddagger$\end{tabular}\thanks{This study was carried out within the LICAM project – funded by European Union – Next Generation EU  within the PRIN 2022 program (D.D. 104 - 02/02/2022 Ministero dell’Università e della Ricerca). This manuscript reflects only the authors’ views and opinions and the Ministry cannot be considered responsible for them.}}
\address{$^\star$ Department of Information Engineering, CNIT - University of Brescia, Italy\\
$^\dagger$ Department of Informatics, University of Salerno, Italy\\
$^\mathsection$ DSMC, University of Brescia, Italy\\
$^\ddagger$ Department of Computer Science, National Yang Ming Chiao Tung University, Hsinchu,Taiwan}

\begin{document}
%\ninept
%
\maketitle
\begin{abstract}
The incorporation of LiDAR technology into some high-end smartphones has unlocked numerous possibilities across various applications, including photography, image restoration, augmented reality, and more. In this paper, we introduce a novel direction that harnesses LiDAR depth maps to enhance the compression of the corresponding RGB camera images. To the best of our knowledge, this represents the initial exploration in this particular research direction.
Specifically, we propose a Transformer-based learned image compression system capable of achieving variable-rate compression using a single model while utilizing the LiDAR depth map as supplementary information for both the encoding and decoding processes. Experimental results demonstrate that integrating LiDAR yields an average PSNR gain of $0.83$ dB and an average bitrate reduction of $16\%$ as compared to its absence.
\end{abstract}
\section{Introduction}
LiDAR, an acronym for Light Detection and Ranging, is a technology that utilizes laser beams to measure distances and create detailed 3D maps of the surrounding environment~\cite{dong2017lidar}. Apple has recently introduced LiDAR technology into some of its products, including the iPad Pro, which was the first to incorporate this technology in March 2020, followed by the iPhone 12 Pro and iPhone 12 Pro Max (and later editions)~\cite{apple}. The purpose of the LiDAR scanner is depth sensing. It typically emits laser pulses and measures the time they take to bounce back from objects in the environment, thereby generating a precise 3D map of the surroundings through specialized algorithms.

The integration of LiDAR technology in Apple's devices has opened up various possibilities. It significantly enhances augmented reality experiences, making them more immersive and accurate. In fact, the precise depth data obtained from LiDAR allows augmented reality applications to better understand the physical world and seamlessly overlay digital objects onto it. This technology has been employed in a variety of ARKit applications and games, which have seen improved performance and user experience. Furthermore, the LiDAR sensor has also improved photography. Its capabilities extend to enhancing autofocus, especially in low-light conditions, ensuring that users can capture sharp images even in challenging environments. It also contributes to better portrait mode effects and augmented reality photography. Beyond photography and augmented reality, the LiDAR sensor aids in object and scene recognition, with applications in accessibility features designed to assist users with visual impairments in understanding their surroundings.

Of all these applications, none has ever delved into the potential of using the LiDAR depth map to enhance the compression of RGB camera images. The field of image compression has seen continuous research for many decades, and its importance has grown substantially, especially with the widespread use of mobile devices for capturing and sharing images. Lossy image compression, in particular, plays a crucial role in this context, as it efficiently reduces the required storage space and transmission bandwidth, even though it comes at the cost of some degradation in the quality of reconstructed images. Deep learning-based techniques~\cite{balle2016end,balle2018variational,choi2019variable,chen2021end,mentzer2018conditional,mentzer2020high,minnen2020channel} have emerged as front-runners, surpassing traditional codecs such as JPEG~\cite{pennebaker1992jpeg}, JPEG2000~\cite{skodras2001jpeg}, and BPG~\cite{bellardbpg}. Many of these learning-based approaches leverage autoencoder networks to perform nonlinear transformations~\cite{balle2020nonlinear} and optimize the rate-distortion trade-off~\cite{davisson1972rate}, making them promising candidates for the next generation of image compression.

\begin{figure*}[t]
\centering
	{\includegraphics[width=0.9\textwidth]{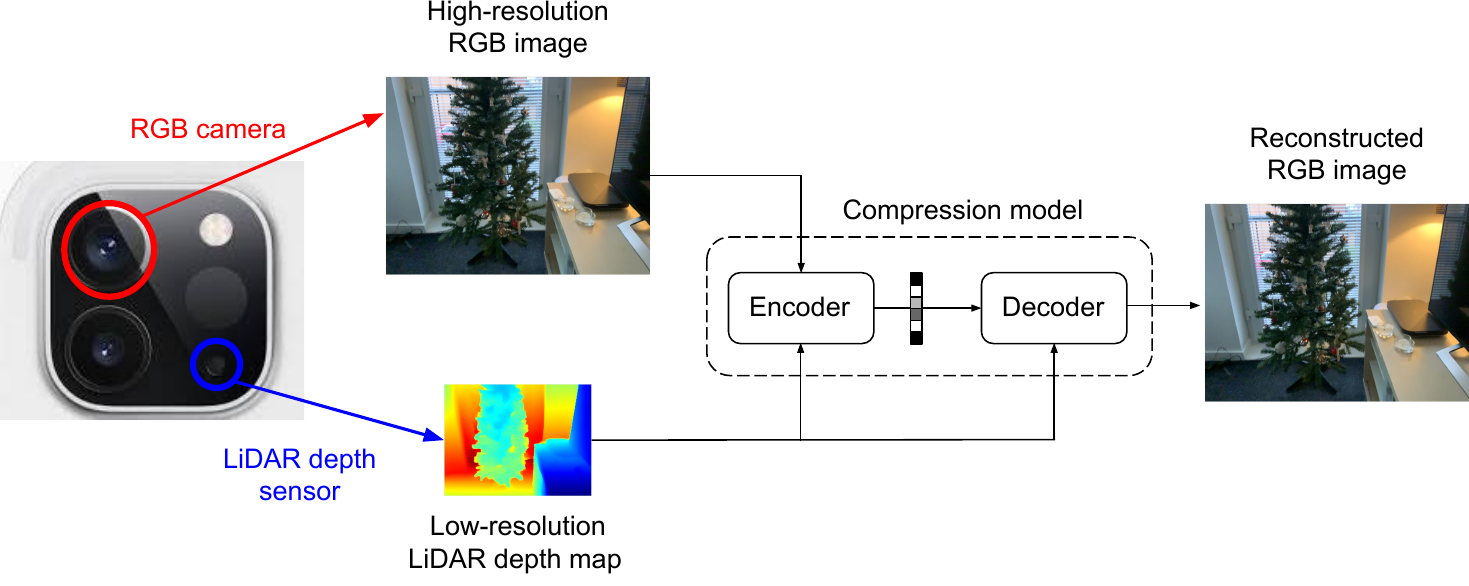}}
\caption{High-level architectural framework. Our LiDAR Depth Map Guided Image Compression Model takes input from both the RGB image and the LiDAR depth data, using the LiDAR map to enhance the compression process of the RGB image. In the considered scenario, the LiDAR map is available at both the encoder and the decoder.}
\label{fig:overview}
\end{figure*}

In this paper, we propose a novel approach where the LiDAR depth map is leveraged to enhance the compression of the corresponding RGB camera image. To the best of our knowledge, this marks the initial exploration of this particular direction. Given the uncharted nature of this field, we opt to begin by considering the simple scenario in which LiDAR information is available both at the encoder and the decoder.
%Thus, our aim is to store an RGB image on a device while also having access to the associated LiDAR map.
To evaluate our approach, we choose to adopt the Swin-Transformer-based image compression system proposed in~\cite{kao2023transformer} as the reference architecture for our experiments. In fact, in the realm of learned image compression systems, Transformers have emerged as a compelling alternative to convolutional neural networks (CNNs). Their attention-based convolution, in combination with the shifted-windowing technique, offers a balance of superior compression performance and reduced computational requirements. Thus, we opt to incorporate LiDAR information into this architecture, which stands as one of the current state-of-the-art methods for learned image compression. Experimental results show that the integration of LiDAR technology results in an average improvement of $0.83$ dB in PSNR-RGB performance and an average reduction of $16\%$ in bit rate when compared to its absence.

The rest of the paper is structured as follows. Sec.~\ref{sec:scenario} depicts the specific scenario under investigation. Sec.~\ref{sec:proposed-method} introduces the architecture proposed for incorporating LiDAR information, and Sec.~\ref{sec:exp} shows the experimental results. Finally, Sec.~\ref{sec:conc} provides the concluding remarks.

\section{Examined Scenario}
\label{sec:scenario}
\noindent
%Incorporating LiDAR technology into Apple's devices enables the simultaneous capture of photos with the traditional RGB camera and the associated depth map through the LiDAR camera. The aim of this paper is to explore the potential of utilizing LiDAR data to enhance the compression of RGB images. Given the uncharted nature of this field, we initially position ourselves in the simplest scenario to verify whether the LiDAR map can truly assist in the compression of an RGB image.
Let us consider a simple scenario to assess the effectiveness of using a LiDAR map in compressing an RGB image, that is, we assume that the LiDAR map is available at both the encoder and the decoder, as illustrated in Fig.~\ref{fig:overview}. In light of this scenario, we aim to investigate two distinct cases.

As a first case, we consider to have the LiDAR map available in its original \textbf{uncompressed} format. Thus, we examine the situation where the RGB image and the corresponding LiDAR map are simultaneously acquired from the device, and the uncompressed LiDAR map is stored without compression on the acquisition device, as it will be utilized for the various imaging manipulations previously mentioned. Then, the uncompressed LiDAR data will assist in both encoding the RGB image, which will be then locally stored on the device, and decoding the reconstructed version. In essence, this approach capitalizes on the presence of the LiDAR map on the device to enhance the compression of the RGB image. In this context, we will not consider the rate cost of the LiDAR map, as its presence on the device is independent of the RGB image compression goal. Of course, this initial case is confined to an encoding and decoding process performed locally on the device.

\begin{figure*}[t]
\centering
	{\includegraphics[width=0.95\textwidth]{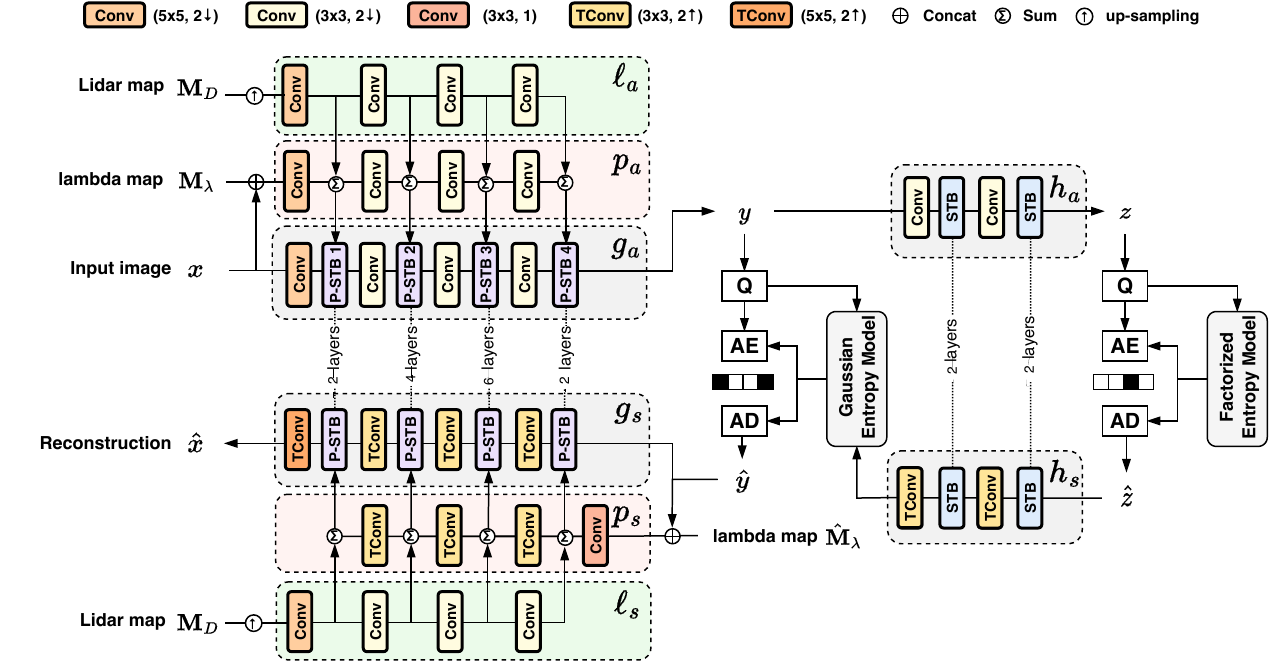}}
\caption{The network architecture of the proposed LiDAR depth map guided image compression model. For variable rate control, the prompt generation networks $p_a$ and $p_s$ produce prompt tokens for the encoder $g_a$ and decoder $g_s$, respectively. Two additional prompt generation networks, denoted as $l_a$ and $l_s$, receive the LiDAR map as input and generate prompt tokens, which are subsequently combined at the convolutional layers of the prompt networks $p_a$ and $p_s$, respectively.}
\label{fig:arch}
\end{figure*}

Secondly, we examine the case in which the RGB image must be transmitted to a second device. As transmitting the uncompressed LiDAR map solely for image replication on a remote device would be impractical due to the high transmission cost, we explore the impact of employing a \textbf{compressed} LiDAR map within the considered image compression framework and its associated rate cost. Thus, in this context, the LiDAR map is first compressed independently, and then the compressed version is used to assist the encoding on the first device and the decoding on the second device of the RGB image. It is important to highlight that, for the image encoding process, we would have the option to employ the uncompressed LiDAR map, assuming it has been stored on the acquisition device without compression (or at a very high quality). However, we opt to use the compressed LiDAR map for assisting the image encoding as well, ensuring uniformity in the LiDAR data used for both image encoding and decoding. It is also important to note that incorporating a compressed depth map into the bitstream typically results in a marginal increase in coding cost, even due to the considerably lower spatial resolution of the LiDAR map compared to the RGB image (see Sec.~\ref{subsec:dataset}). A similar framework is discussed in~\cite{agustsson2019generative}, where indeed a semantic/instance label map of the original image is utilized as side information at both the encoder and decoder for extreme low rate coding.

We would like to remark that in the case of transmitting the LiDAR map alongside, an effective solution might involve jointly compressing both the RGB image and the LiDAR depth map. This approach would entail designing a model that takes both the RGB image and LiDAR map as inputs, encoding both of them, and producing both their reconstructed versions as outputs. Alternatively, one could envision a model that uses the LiDAR map exclusively at the encoder side, without involving it in the decoding process. In fact, despite the established principle derived from Berger's result, indicating that side information, when solely available at the encoder and not influencing the distortion measure, does not offer benefits~\cite{berger1971}, one could explore whether incorporating LiDAR data at the encoder could improve the convergence of an end-to-end learning model and potentially result in a more efficient data compression model. We leave these potential investigations for future works.

\section{Proposed method}
\label{sec:proposed-method}
\noindent
Following the approach presented in~\cite{kao2023transformer}, we propose to adopt their Swin-Transformer-based image compression system as our reference architecture. The structure of our LiDAR depth map guided image compression model is shown in Fig.~\ref{fig:arch}. 

\subsection{Background: reference architecture}
\label{subsec:ref-arch}
\noindent
The system introduced in~\cite{kao2023transformer} is an evolution of TIC (Transformer-based Image Compression~\cite{lu2022transformer, lu2021transformer}), but it omits the inclusion of a context model for entropy coding. The core components of this system consist of a main autoencoder, denoted as $g_a$ and $g_s$, and a hyperprior autoencoder, represented as $h_a$ and $h_s$, which incorporate Swin-Transformer blocks (STB) interspersed with convolutional layers. More details about STBs are presented in~\cite{lu2021transformer}.

As usually performed in learned end-to-end image compression frameworks, the analysis transform $g_a$ encodes an image $x \in \mathbb{R}^{3 \times H \times W}$ into its latent representation $y \in \mathbb{R}^{\frac{H}{16} \times \frac{W}{16} \times 192}$. This latent representation $y$ undergoes uniform quantization as $\hat{y}$ and is further encoded into a bitstream using a learned prior distribution. On the decoder side, the bitstream is entropy decoded to reconstruct $\hat{x} \in \mathbb{R}^{3 \times H \times W}$ via the synthesis transform $g_s$. The prior distribution significantly impacts the required number of bits to signal the quantized latent $\hat{y}$. Therefore, it is modeled in a content-adaptive manner by a hyperprior autoencoder, consisting of a hyperprior analysis transform $h_a$ and a hyperprior synthesis transform $h_s$. As illustrated in Fig.~\ref{fig:arch}, $h_a$ converts the latent $y$ into $z \in \mathbb{R}^{\frac{H}{64} \times \frac{W}{64} \times 192}$, typically representing a small portion of the compressed bitstream. Its quantized version $\hat{z}$ is ultimately decoded from the bitstream through $h_s$. A factorized entropy method is applied for $\hat{z}$, while a gaussian entropy model is used to characterize $\hat{y}$ for probability estimation (see~\cite{balle2018variational}).

In~\cite{kao2023transformer}, to encode the input image $x$, the encoder requires two additional inputs: a lambda map $M_\lambda \in \mathbb{R}^{1 \times H \times W}$ and an ROI mask. The lambda map $M_\lambda$ is a uniform map with a parameter $m_\lambda \in [0, 1]$, which regulates the bit rate of the compressed bitstream. The ROI mask specifies the spatial importance of individual pixels in the image. Both of these inputs serve as conditioning signals that are used to generate prompt tokens for adapting the primary encoder $g_a$. Likewise, the decoder $g_s$ is adapted by receiving inputs in the form of the quantized latent $\hat{y}$ and a downscaled lambda map $\hat{M}_\lambda \in \mathbb{R}^{1 \times \frac{H}{16} \times \frac{W}{16}}$, which matches the spatial resolution of $\hat{y}$.

These prompts enable variable-rate and ROI coding and draw inspiration from~\cite{jia2022visual}. The outcome is a modified Swin-Transformer block termed prompted Swin-Transformer block (P-STB). The generation of these learned prompts involves two networks: $p_a$ and $p_s$, conditioning the encoder $g_a$ and decoder $g_s$, respectively. The $p_a$ network comprises several convolutional layers matching those of the encoder $g_a$ and takes as input the concatenation of the ROI mask, the lambda map $M_\lambda$, and the image $x$. The feature maps generated by $p_a$ are channeled into the respective P-STBs to produce prompt tokens, which subsequently interact with image tokens. On the other hand, $p_s$ follows a similar architectural pattern, replacing the convolutional layers with transposed convolutional layers for upsampling. More details can be found in~\cite{kao2023transformer}.

\subsection{LiDAR Depth Map Guided Image Compression Model}
\label{subsec:lidar-depth-map-model}
\noindent
Fig.~\ref{fig:arch} depicts our image compression model guided by LiDAR depth maps.  It draws inspiration from~\cite{kao2023transformer}. However, two significant distinctions set our approach apart from~\cite{kao2023transformer}. Firstly, our method does not incorporate the ROI mask, as it plays no role in our model, in contrast to the approach outlined in~\cite{kao2023transformer}, which proposes a ROI-based coding. Secondly, the distinctive feature of our model lies in the inclusion of two supplementary prompt generation networks, referred to as $l_a$ and $l_s$ situated at the encoder and decoder sides, respectively. Both of them receive as input the LiDAR depth map $M_D \in \mathbb{R}^{1 \times H \times W}$ and generate prompt tokens, which are then added to the prompt tokens of $p_a$ and $p_s$, respectively.
Note that the LiDAR depth map $M_D$ undergoes an initial upsampling process to align its size with that of the RGB image. This is necessary because the resolution of LiDAR sensors is typically smaller ($256 \times 192$ for Apple's devices), compared to the RGB camera.%These prompts allow the model to leverage the LiDAR information effectively for both the encoding and decoding processes.

We would like to remark that in ~\cite{kao2023transformer}, the ROI map information is concatenated with the lambda map $M_\lambda$ (only at the encoder side). However, we found that concatenating the LiDAR depth map with $M_\lambda$ does not allow the model to significantly exploit the depth information. Instead, incorporating the specific prompts $l_s$ and $l_a$ at both the encoder and decoder sides, which receive both $M_D$, enables the system to effectively leverage the LiDAR depth information (as we will show in the next section).

In training, the loss function is formulated to make the model responsive to the rate parameter $m_\lambda$, without incorporating any LiDAR information into the loss function itself. Thus, it is defined as a weighted combination of both distortion and bit rate components:

\begin{equation}
    J_{rd}(x) = \lambda \sum_{i=1}^N \frac{(x_i - \hat{x}_i)^2}{N} + R,
\end{equation}
where $x_i$ and $\hat{x}_i$ are the $i$-th pixel in the original and reconstructed images, respectively, $N$ is the total number of pixels in the image, $R$ is the bit rate and denotes the number of bits-per-pixel needed to represent the latent representations $\hat{y}$ and $\hat{z}$, and $\lambda = f(m_\lambda) = (\lambda_{\text{max}})^{m_{\lambda}} (\lambda_{\text{min}})^{1-m_{\lambda}}$ is a Lagrange multiplier, which is a function of the rate parameter $m_\lambda$, with $\lambda_\text{max}$ and $\lambda_\text{min}$ being the highest and lowest $\lambda$, respectively.

\begin{figure*}[]
    \centering
    \subfloat[][Average RD curves (PSNR).]
    {\includegraphics[width = 1\columnwidth]{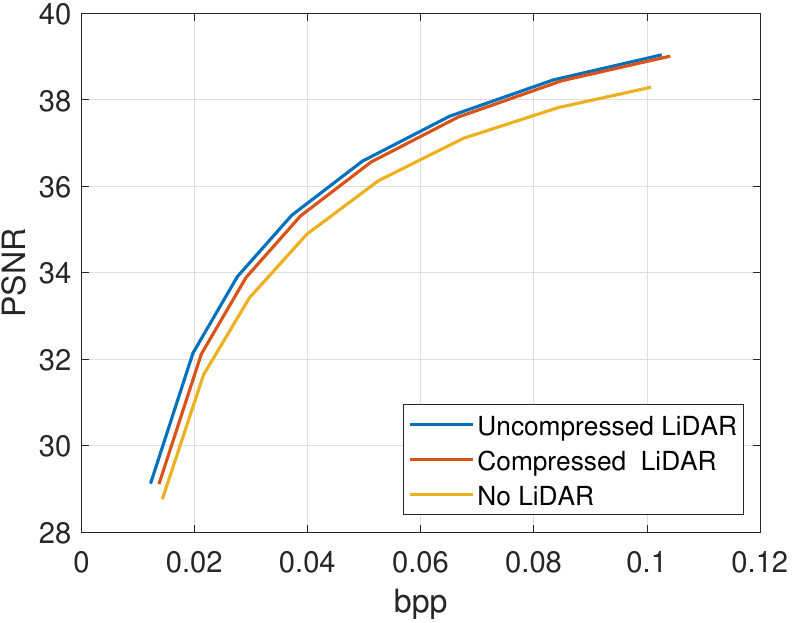}\label{fig:psnr}}\:\:
    \subfloat[][Average RD curves (SSIM).]
    {\includegraphics[width = 1\columnwidth]{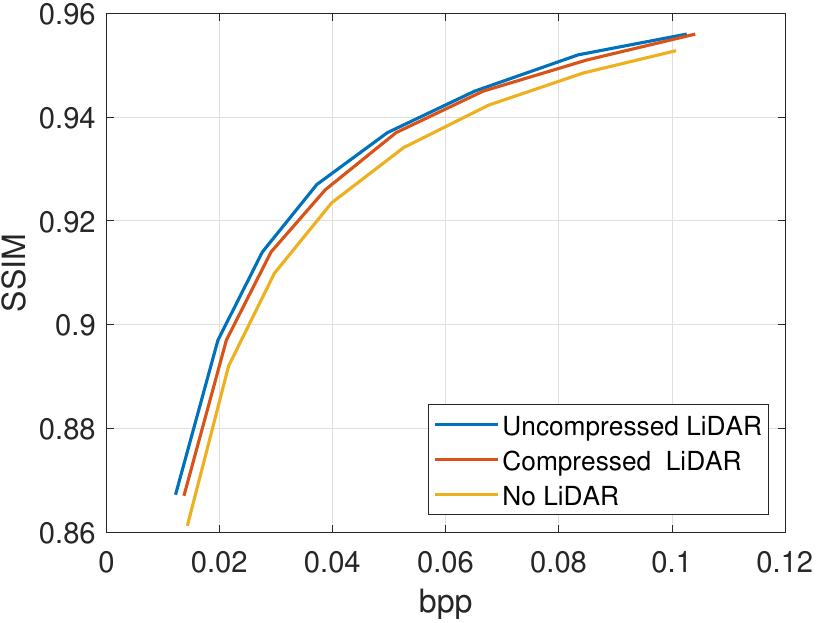}\label{fig:ssim}}
    \caption{Comparison of the model performance with and without LiDAR in terms of average RD curves, evaluating PSNR (a) and SSIM (b) as distortion metrics. The blue curve corresponds to the model's performance when given the uncompressed LiDAR map, while the red curve represents the model's performance when supplied with the compressed LiDAR map, which includes the cost of transmitting LiDAR information. The results are evaluated on the $5,999$ images in the test set.}
    \label{fig:results}
\end{figure*}

\section{Experiments}
\label{sec:exp}
\noindent
\subsection{Dataset}
\label{subsec:dataset}
\noindent
The choice of datasets is critical to ensure they accurately represent the smartphone-based sensor fusion scenario. This poses a challenge because smartphones equipped with both cameras and LiDAR sensors have only become available recently, and traditional RGB-D datasets were primarily designed for dedicated devices like the Microsoft Kinect. Among the existing datasets, ARKitScenes~\cite{baruch2021arkitscenes}, constructed using the built-in cameras and LiDAR sensors in Apple smartphones and tablets, is the only one that aligns with our scope.

The dataset consists of indoor scene images captured using the 2020 iPad Pro, which was utilized to collect both RGB images from the RGB camera and dense depth maps from the LiDAR scanner. The dataset comprises a total of $44,599$ acquisitions, each one including both the low-resolution and high-resolution RGB images ($256 \times 192$ and $1920 \times 1440$ pixels, respectively), and the corresponding low-resolution depth maps ($256 \times 192$ pixels). To work with high-resolution data, we performed upsampling on the low-resolution depth maps, increasing their resolution by a factor of $7.5$. In line with the existing division in the ARKitScenes dataset, we partitioned the dataset into a training set, which includes $39,000$ images, and a test set, consisting of $5,599$ images.

\subsection{Results}
\noindent
We conduct a single-stage training of our model using the ARKitScenes dataset's training set. During each training iteration, input images are randomly cropped to a size of $256 \times 256$ and the uncompressed LiDAR maps undergo cropping in corresponding positions. Our entire model is trained jointly for variable rate coding by uniformly sampling $m_\lambda$ from $0$ to $1$. To ensure fairness in comparison, we adhere to the same training procedure for the model without LiDAR prompts; that is, the model presented in~\cite{kao2023transformer} when considering uniform ROI maps.

Fig.~\ref{fig:results} shows the rate-distortion (RD) curves, analyzing both PSNR-RGB (used in the training loss function) and SSIM, shown separately in Fig.~\ref{fig:psnr} and Fig.~\ref{fig:ssim}. The blue curve represents the model adopting the uncompressed LiDAR for both the encoding and decoding stages. As it can be observed, this model consistently outperforms the model without LiDAR for both metrics.  In terms of Bj\o ntegaard's metric, integrating uncompressed LiDAR results in an average PSNR gain of $0.83$ dB and an average bit rate reduction of $16\%$ compared to its absence.

Furthermore, the red curve corresponds to the performance of our model when considering the case in which the uncompressed LiDAR data is unavailable, such as during transmission. In this context, the LiDAR information needs to be compressed separately. In our experiments, it undergoes separate compression using our model, where we specifically provide zero matrices as input to the prompts $l_a$ and $l_s$ and set $m_{\lambda} = 1$ to maximize the quality of the compressed LiDAR map.  Then, the compressed LiDAR map is used for assisting the RGB image encoding and decoding. Of course, the final bitstream will consist of both the latent spaces of the RGB image and the LiDAR map. Interestingly, the resulting performance is comparable to that achieved by the model using the original LiDAR map, both in terms of PSNR and SSIM. In fact, the increase in coding cost when encoding the LiDAR depth map is only marginal. On average, it accounts for about 9\% of the total cost, which includes both the RGB image and the LiDAR map. This is due to the significantly lower spatial resolution of the LiDAR map compared to the RGB image. Moreover, using the compressed LiDAR map in place of the original one does not significantly affect the compression process. Specifically, in terms of Bj\o ntegaard's metric, integrating compressed LiDAR results in an average PSNR gain of $0.57$ dB and an average bit rate reduction of $10\%$ compared to its absence (see Tab.~\ref{tab:bjo} for recap). Notably, we remark that we have not retrained the model using the compressed LiDAR maps.

\begin{table}[t]
\begin{tabular}{@{}ccc@{}}
\toprule
                       & Uncompressed LiDAR & Compressed LiDAR \\ \midrule
$\Delta$ PSNR {[}dB{]} & 0.83               & 0.57             \\
$\Delta$ rate {[}\%{]} & -16                & -10              \\ \bottomrule
\end{tabular}
\caption{\textit{BD} rate associated to Fig.~\ref{fig:psnr}. The baseline performance is set with the model that does not use the LiDAR map (no LiDAR).}
\label{tab:bjo}
\end{table}

\begin{figure*}[]
\centering
\subfloat[][Original image.]
	 {\setcounter{subfigure}{1}\includegraphics[width = 0.18\textwidth]{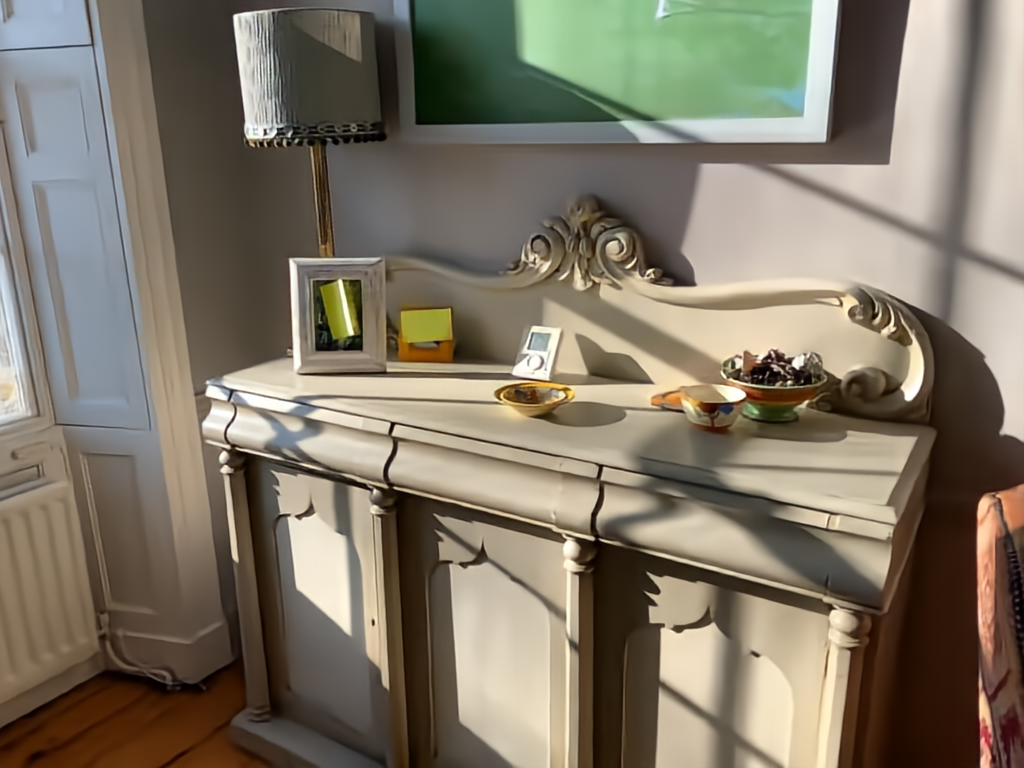}\label{fig:img1}}\
\subfloat[][Rec. (LiDAR).\\Bpp: 0.15 - PSNR: 33.09.]
	 {\setcounter{subfigure}{3}\includegraphics[width = 0.18\textwidth]{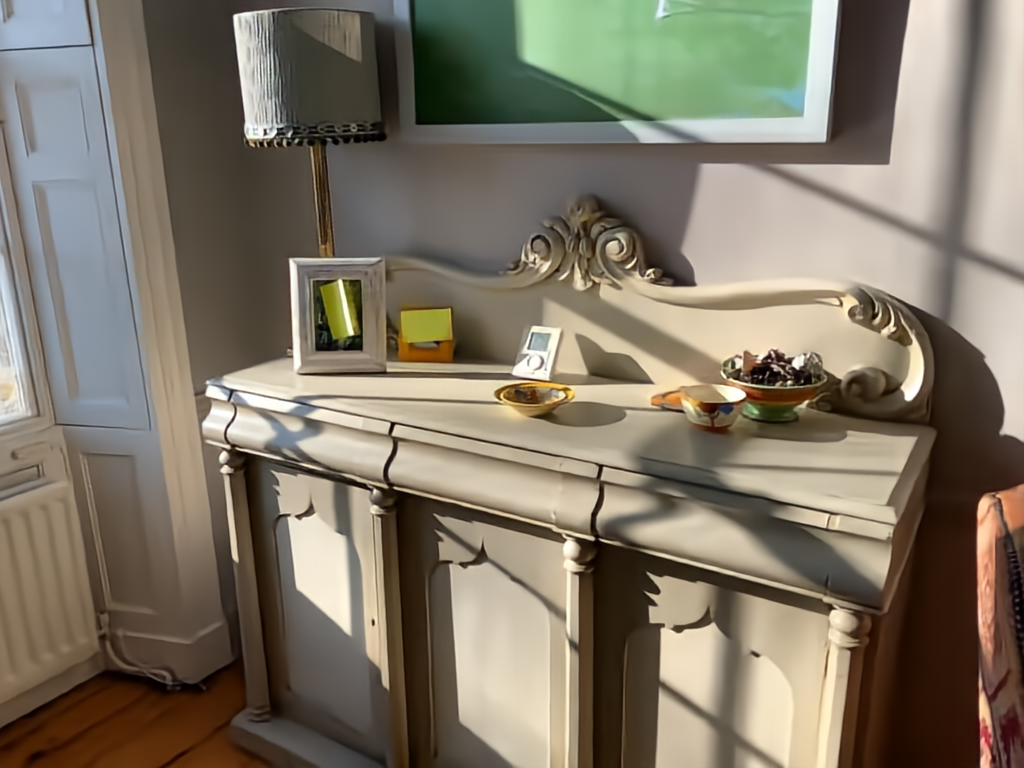}\label{fig:img1-lidar}}\
\subfloat[][Rec. (no LiDAR).\\Bpp: 0.16 - PSNR: 32.38.]
	 {\setcounter{subfigure}{5}\includegraphics[width = 0.18\textwidth]{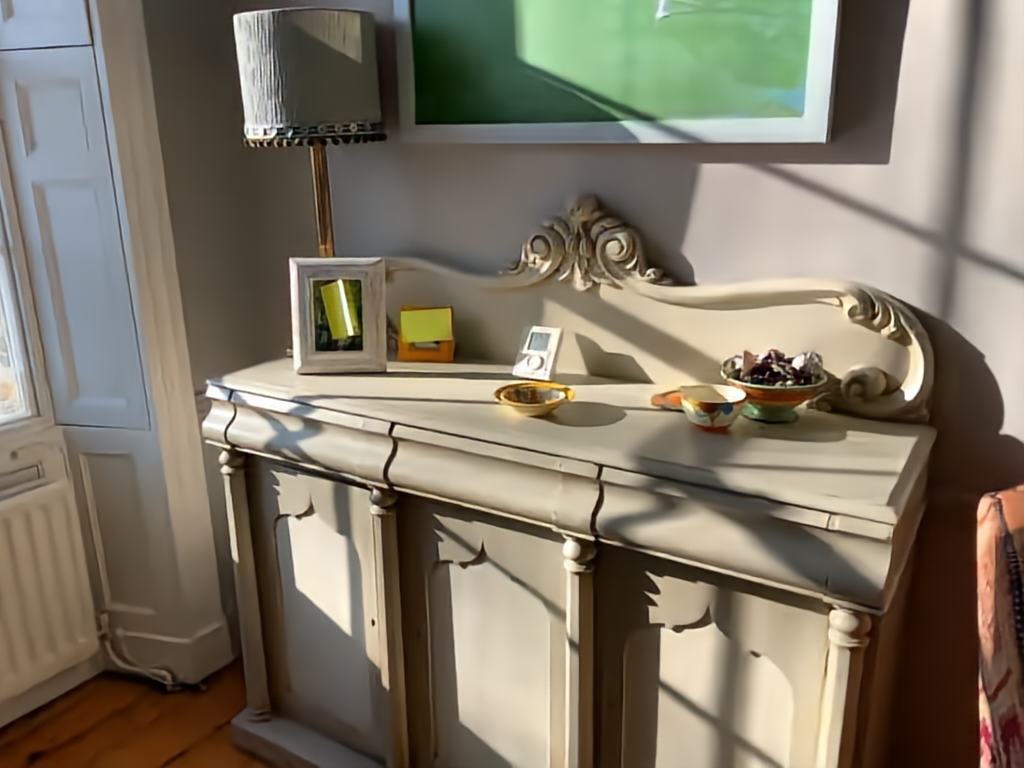}\label{fig:img1-no-lidar}}\
\subfloat[][Rec. (random LiDAR).\\Bpp: 0.15 - PSNR: 30.69.]
	 {\setcounter{subfigure}{7}\includegraphics[width = 0.18\textwidth]{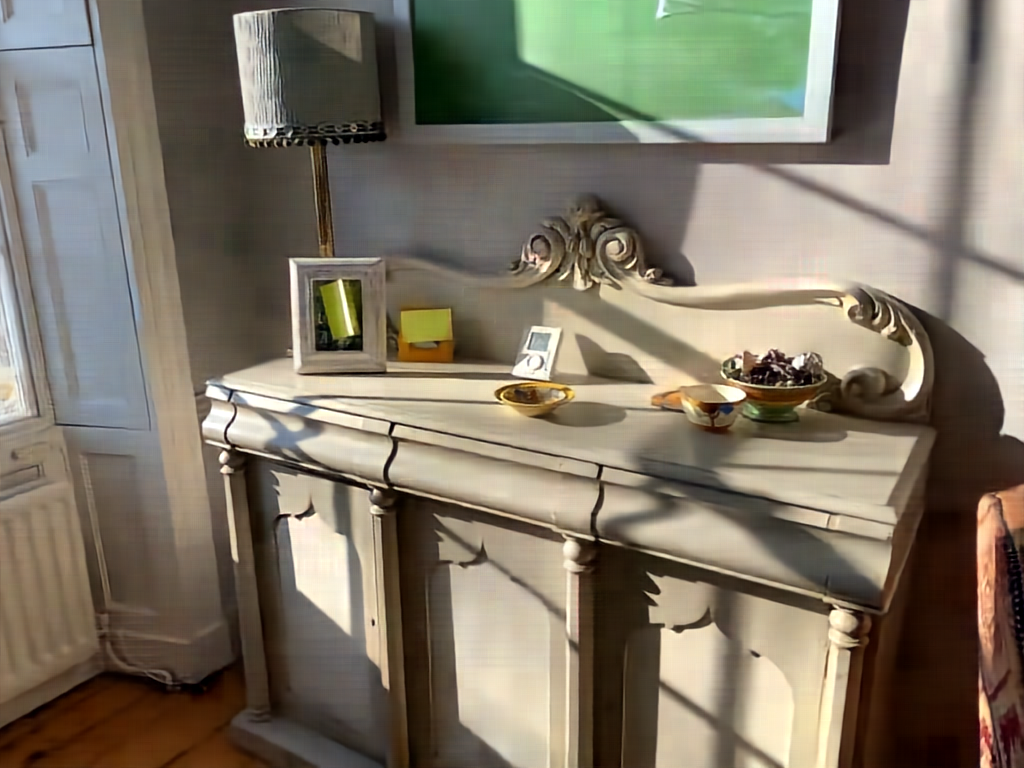}\label{fig:img1-random-lidar}}\
\subfloat[][LiDAR depth map.]
	 {\setcounter{subfigure}{9}\includegraphics[width = 0.18\textwidth]{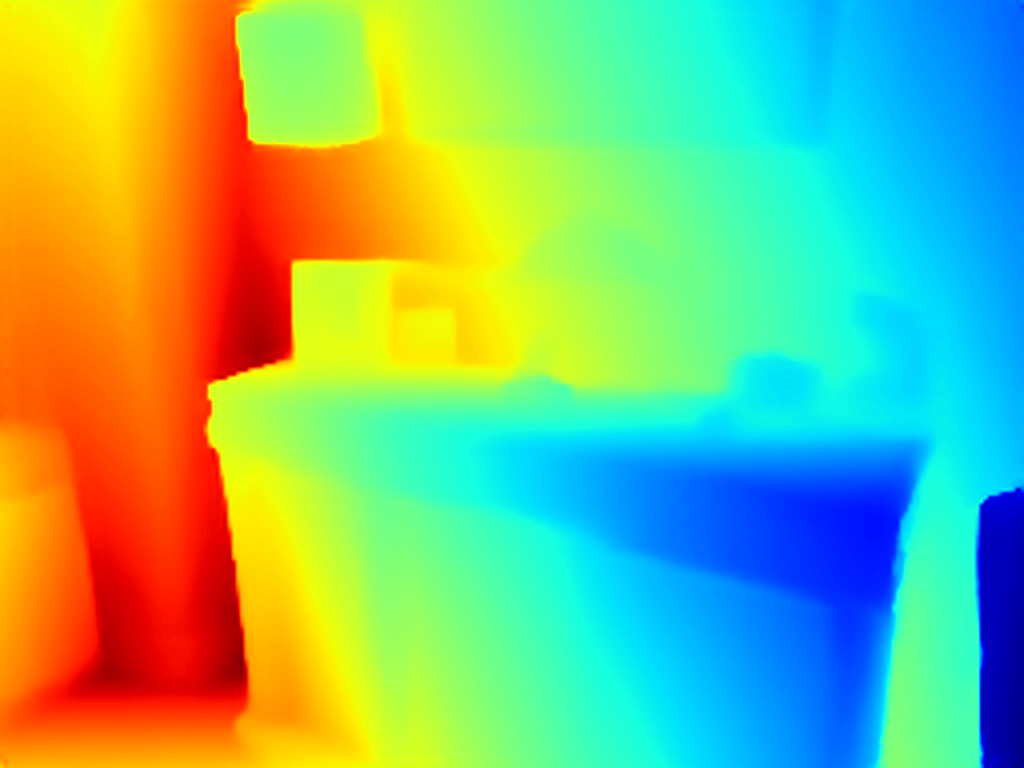}\label{fig:depthmap1}}\\
\subfloat[][Original image.]
	 {\setcounter{subfigure}{2} \includegraphics[width = 0.18\textwidth]{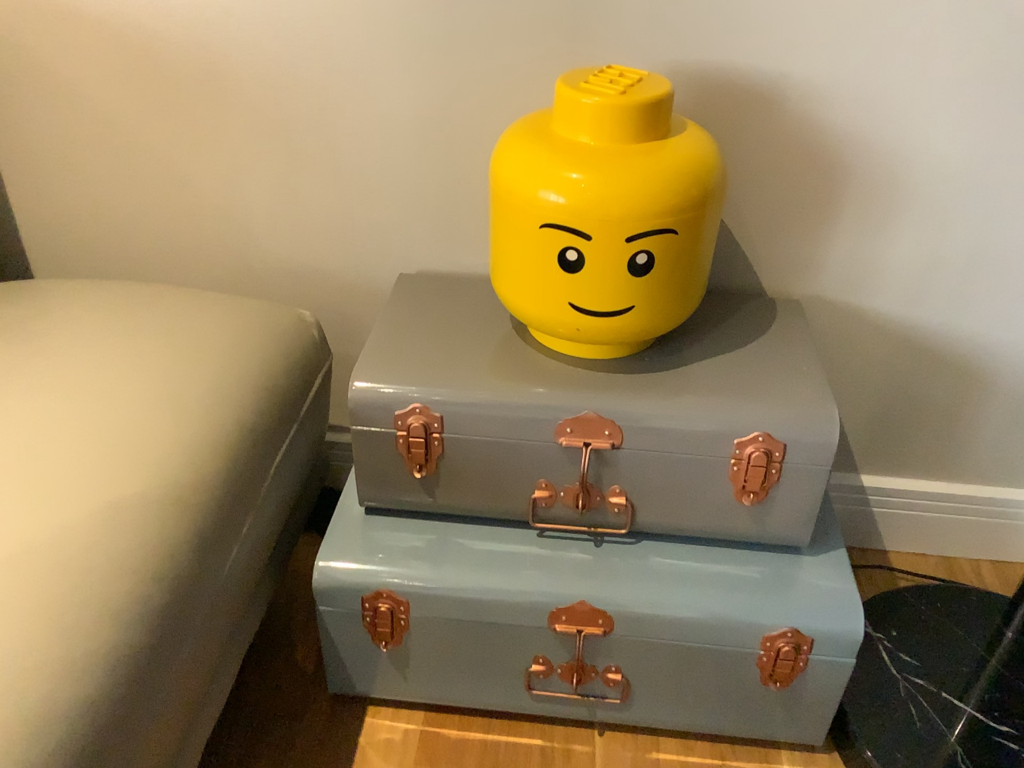}\label{fig:img2}}\
\subfloat[][Rec. (LiDAR).\\Bpp: 0.03 - PSNR: 33.03.]
	 {\setcounter{subfigure}{4}\includegraphics[width = 0.18\textwidth]{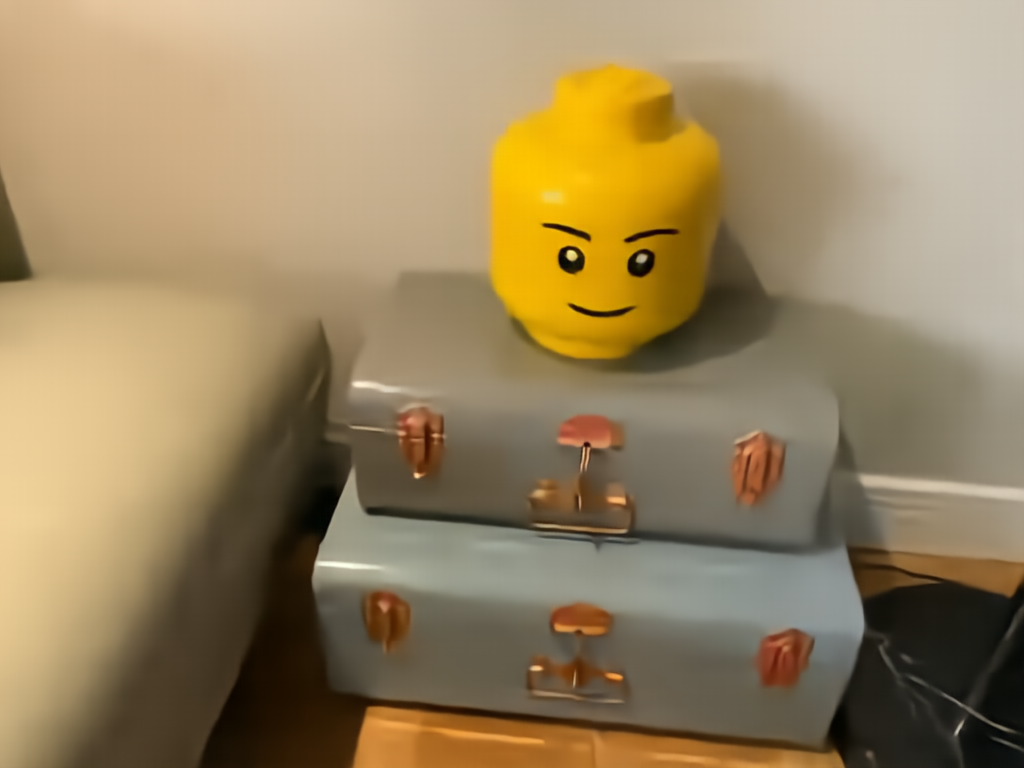}\label{fig:img2-lidar}}\
\subfloat[][Rec. (no LiDAR).\\Bpp: 0.03 - PSNR: 32.09.]
	 {\setcounter{subfigure}{6}\includegraphics[width = 0.18\textwidth]{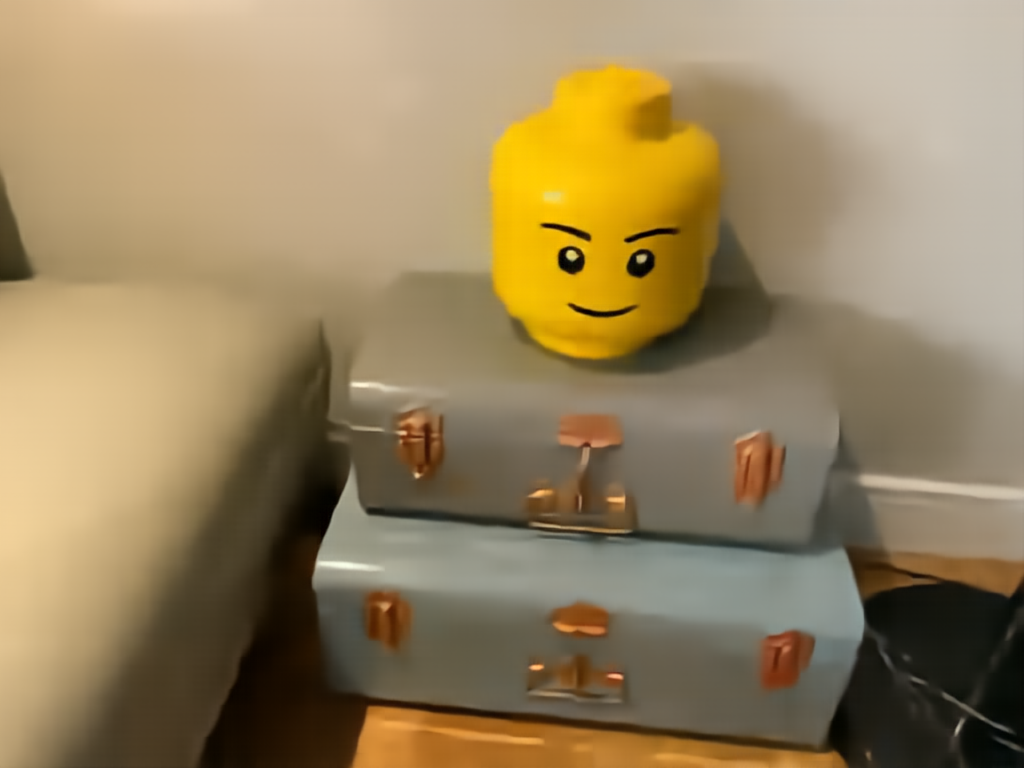}\label{fig:img2-no-lidar}}\
\subfloat[][Rec. (random LiDAR).\\Bpp: 0.15 - PSNR: 29.71.]
	 {\setcounter{subfigure}{8}\includegraphics[width = 0.18\textwidth]{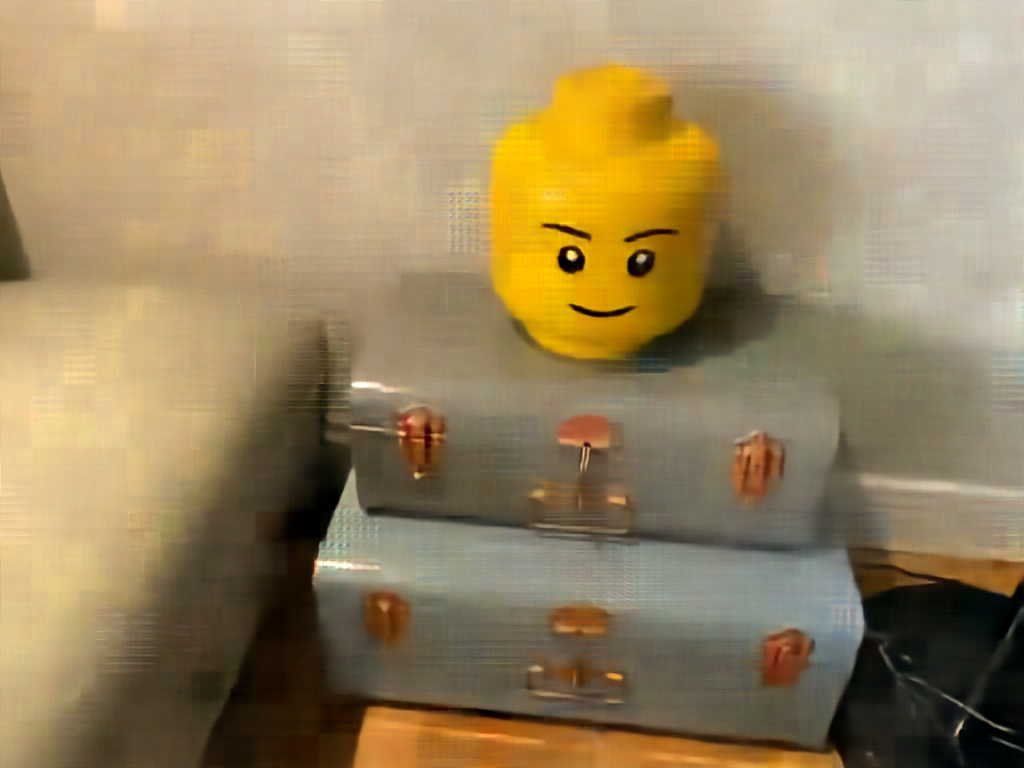}\label{fig:img2-random-lidar}}\
\subfloat[][LiDAR depth map.]
	 {\setcounter{subfigure}{10}\includegraphics[width = 0.18\textwidth]{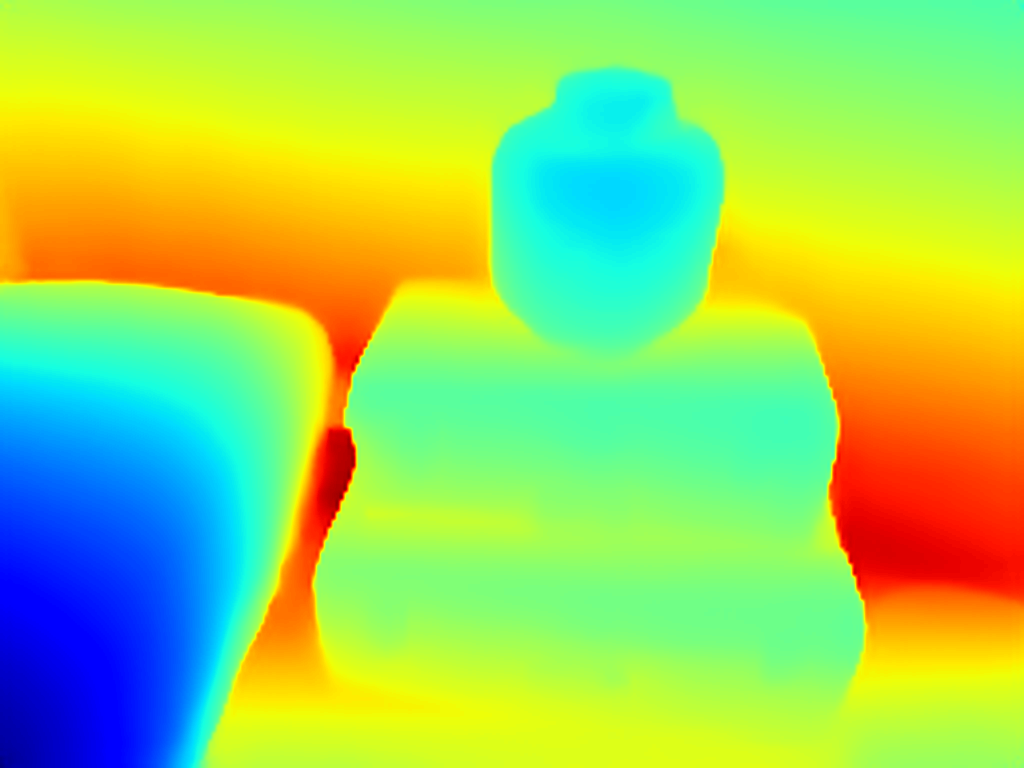}\label{fig:depthmap2}} 
\caption{Visual results for two images from the test set are displayed in (a)-(b). The reconstructed images using LiDAR as side information are depicted in (c)-(d), while those reconstructed without LiDAR are shown in (e)-(f). The reconstructed images using a random map as side information are seen in (g)-(h). Corresponding LiDAR maps are represented in (i)-(j). %It is recommended to zoom in for a better appreciation of the visual differences.
}
\label{fig:visual-results}
\end{figure*}

To better appreciate the visual quality, Fig.~\ref{fig:visual-results} displays the reconstructed images for two samples from the test set (Figs.~\ref{fig:visual-results}a and \ref{fig:visual-results}b). Specifically, it showcases the results using our model with LiDAR input (Figs.~\ref{fig:visual-results}c and \ref{fig:visual-results}d) and the model without LiDAR (Figs.~\ref{fig:visual-results}e and \ref{fig:visual-results}f). Additionally, to highlight the effectiveness of LiDAR integration within our model, Figs.~\ref{fig:visual-results}g and \ref{fig:visual-results}h display the reconstructed images from our model when a random map serves as the side information. Evidently, there is a notable decrease in visual quality in these instances, indicating the significant impact of LiDAR data on the compression process.

\section{Conclusion}
\label{sec:conc}
\noindent
In this paper, we have shown that integrating the LiDAR depth map within the compression process of an RGB camera image results in substantial performance improvements. As the use of LiDAR in mobile devices continues to grow, our work can serve as a starting point for a new research direction. There are some aspects that deserve further exploration in our study: (i) extending the experiments to include additional and more variegated RGB-LiDAR datasets, which are expected to increase considering the growing prevalence of LiDAR technology; (ii) investigating the universality of the method by integrating the LiDAR map into various image compression frameworks and assessing its efficiency; (iii) while our findings demonstrate that utilizing a decoded LiDAR map at the decoder side does not compromise the performance compared to the model using the original map, we believe that employing a combined compression approach for both the image and LiDAR map could potentially yield even greater enhancements. We leave this possibility for exploration in future research endeavors.

\bibliographystyle{IEEEbib}
\bibliography{refs}

\end{document}